\documentclass[12pt]{article}
\usepackage{amssymb,epsf}
\usepackage{graphicx}
\newcommand{\beq}{\begin{equation}}
\newcommand{\eeq}{\end{equation}}
\newcommand{\bea}{\begin{eqnarray}}
\newcommand{\eea}{\end{eqnarray}}
\newcommand{\bdm}{\begin{displaymath}}
\newcommand{\edm}{\end{displaymath}}

\newcommand{\ra}{\rightarrow}

\def\<{\langle}
\def\>{\rangle}
\def\slashed#1{{\ooalign{\hfil\hfil/\hfil\cr $#1$}}}

\def\a{\alpha}
\def\b{\beta}

\def\d{\delta}
\def\e{\epsilon}           % Also, \varepsilon
\def\g{\gamma}

\def\p{\pi}                % Also, \varpi
\def\th{\theta}                   %     \vartheta
\def\t{\tau}

\def\D{\Delta}

\def\O{\Omega}

\def\TH{\Theta}

\def\cn{{\cal N}}

\newcommand{\Tr}{\mathop{\rm Tr}\nolimits}

\def \rr {{\mathbb R}}

\def\slashed#1{{\ooalign{\hfil\hfil/\hfil\cr $#1$}}}

\def\N{{\mathcal{N}}}

\begin{document}
\baselineskip=15.5pt
\pagestyle{plain}
\setcounter{page}{1}
%--------+---------+---------+---------+---------+---------+---------+
%Body

%\begin{document}

\begin{flushright}
{\tt hep-th/0605120}
\end{flushright}

\vskip 2cm

\begin{center}
{\Large \bf Chiral Transition of N=4 Super Yang-Mills with Flavor on a 3-Sphere}
\vskip 1cm

{\bf Andreas Karch and Andy O'Bannon} \\
\vskip 0.5cm
{\it Department of Physics, University of Washington, \\
Seattle, WA 98195-1560 \\}
{\tt  E-mail: karch@phys.washington.edu, ahob@u.washington.edu} \\
\medskip

\end{center}

\vskip1cm

\begin{center}
{\bf Abstract}
\end{center}
\medskip

We use the AdS/CFT correspondence to perform a numerical study of a
phase transition in strongly-coupled large-$N_c$ $\mathcal{N} = 4$
Super-Yang-Mills theory on a 3-sphere coupled to a finite number
$N_f$ of massive $\mathcal{N} = 2$ hypermultiplets in the
fundamental representation of the gauge group. The gravity dual
system is a number $N_f$ of probe D7-branes embedded in $AdS_5
\times S^5$. We draw the phase diagram for this theory in the plane
of hypermultiplet mass versus temperature and identify for
temperatures above the Hawking-Page deconfinement temperature a
first-order phase transition line across which the chiral condensate
jumps discontinuously.

\newpage

\section{Introduction}
\setcounter{equation}{0}

The AdS/CFT correspondence \cite{Maldacena:1997re} equates the low-energy
effective theory of string theory, supergravity (SUGRA), on the background
$AdS_5 \times S^5$ with strongly-coupled $\mathcal{N} = 4$ $SU(N_c)$
super-Yang-Mills theory (SYM) in the large-$N_c$ limit living, in some
sense, on the boundary of $AdS_5$. At finite temperature this
four-dimensional boundary is $S^1 \times S^3$. This field theory contains
only adjoint fields and exhibits a first-order deconfinement transition at
the Hawking-Page temperature, corresponding to AdS-Schwarzschild black
hole condensation \cite{Witten:1998zw}. The thermodynamics of this SYM theory has
been studied in great detail both at strong coupling via AdS/CFT and at
zero \cite{Sundborg:1999ue} and weak coupling \cite{Aharony:2003sx}.

Why consider a theory on a compact space? For this conformal theory, the
deconfinement transition cannot occur in infinite volume, i.e. on
spacetime $S^1 \times \rr^3$ because with no scale nothing can set a transition
temperature. The $S^3$ introduces a new scale, and indeed the
deconfinement transition occurs at a temperature of order the inverse
$S^3$ radius \cite{Aharony:2003sx}. On the gravity side, this is the statement
that the Hawking-Page transition is only apparent in global AdS
coordinates. Poincare patch coordinates are limited to the
high-temperature black hole solution, corresponding to the deconfined
phase in the SYM theory. Of course, no phase transitions can occur with a
finite number of degrees of freedom, which then motivates the 't Hooft limit.

In large-$N_c$ Yang-Mills, or more generally in gauge theories with only
adjoint fields, the deconfinement transition comes from order $N_{c}^2$
dynamics. One way to identify the deconfinement transition is to ask whether the order $N_{c}^2$
contribution to the free energy is zero or not. In pure Yang-Mills, for
example, at sufficiently low temperature this contribution to the free
energy will be zero, as the degrees of freedom are color singlet glueballs
of whom there are order $N_{c}^0$ (in the large-$N_c$ book-keeping),
while at sufficiently high temperature it will be nonzero, indicating that
the contributing degrees are freedom are deconfined gluons of which there
are order $N_{c}^2$. This is what the word ``deconfinement" means in the
$\N = 4$ SYM theory on a 3-sphere studied at strong coupling using
AdS/CFT.

Introducing fundamental fields into this field theory requires introducing
an open string sector. In reference \cite{Karch:2002sh} D7 flavor branes were introduced to the usual AdS/CFT construction, which equates the field theory on a stack of coincident D3-branes with gravity in the geometry supported
by the 3-branes. Keeping the number $N_f$ of D7's finite while taking the number $N_c$ of D3's to
infinity will produce in the near-horizon limit probe D7-branes embedded in $AdS_5 \times S^5$. These branes will wrap the $AdS_5$ and an $S^3$ inside the $S^5$ factor. The probe branes break half the supersymmetry.

In the field theory this corresponds to adding to the $\N = 4$ theory a number $N_f$ of
$\mathcal{N} = 2$ hypermultiplet fields (henceforth ``hypers") in
the fundamental representation of $SU(N_c)$. The theory will remain conformal in this $N_f / N_c \rightarrow 0$
limit. The hypers can be given a mass by ``ending" the D7-branes at some finite value of the AdS
radius \cite{Karch:2002sh}. Such masses obviously break the conformality but will preserve the $\mathcal{N} = 2$ supersymmetry. On the gravity side this is manifest as supersymmetric probe D7's ending at finite radial coordinate, which
clearly breaks the radial isometry.

The addition of fundamental fields introduces the possibility of a chiral
phase transition analogous to that in quantum chromodynamics (QCD) or at
least large-$N_c$ QCD. By the usual large-$N_c$ counting rules, all
dynamics of fundamental fields is suppressed by a factor of $1/N_c$. For
large-$N_c$ theories, then, the chiral transition involving fundamental
fields is in some sense a small perturbation, suppressed by a power of
$N_c$, on top of the order $N_{c}^2$ dynamics of the deconfinement
transition.

Most of the remainder of  this section, as well as Sections \ref{flavor} and \ref{flat}, contains little new material and is presented to put our work in the proper context. Readers familiar with the topic may want to go directly to Sections \ref{action} and \ref{curved}, where we present our methods and new results. Our main result is figure (\ref{phasediagram}), as summarized in the last three paragraphs of this introduction.

We now review the chiral phase transition in QCD, to compare and
contrast with our $\N=2$ supersymmetric theory on a compact space. We will be careful about two limits: large-$N_c$ and zero quark
mass. In general (any $N_c$ and $N_f$), for massless quarks chiral
$U(N_f)_L \times U(N_f)_R \simeq SU(N_f)_L \times SU(N_f)_R \times
U(1)_B \times U(1)_A$ is an exact symmetry of the Lagrangian. At
zero temperature, $SU(N_f)_L \times SU(N_f)_R$ is spontaneously
broken to the diagonal $SU(N_f)_V$. The chiral symmetry is
restored at sufficiently high temperatures. 
The order parameter for the chiral transition is the quark bilinear
condensate, which is nonzero at low temperatures and zero at high
temperatures, signaling the chiral symmetry restoration. Quark
masses explicitly break the chiral symmetry, but we will follow the
convention of calling the quark bilinear condensate an order
parameter nonetheless.

In QCD with massless quarks, the order of the chiral transition
depends on $N_c$ and $N_f$.  We start with $N_c = 3$. For $N_f \geq
3$ massless quarks, universality arguments indicate that the chiral
transition is first order \cite{Pisarski:1983ms}. For $N_f = 2$, the
chiral transition is believed to be second order
\cite{Wilczek:1993pi, Wilczek:1992sf}\footnote{Lattice simulations have not settled the issue of the order of the transition for $N_f = 2$. As just a small sample: \cite{Karsch:1994hm} found results consistent with a second order transition while more recently \cite{D'Elia:2005sy} found indications of a first order transition}. For $N_f = 1$, things are more subtle. Now the symmetry is simply $U(1)_B \times U(1)_A$ and the question becomes whether the axial symmetry, which is anomalous at $T=0$, is restored as the temperature rises. The instantons responsible for the anomaly are suppressed as $T$ rises, so in fact the axial symmetry is believed to be restored at least in the strict $T = \infty$ limit. As the decrease in the density of instantons is believed to be smooth, no chiral transition exists for $N_f = 1$ \cite{Pisarski:1983ms} as a function of $T$. Turning to the large-$N_c$ limit, for $N_f \geq 2$ the theory has no stable IR fixed point and hence the transition must be first order \cite{Pisarski:1983ms}. At large $N_c$, the axial anomaly is suppressed and to our knowledge it is not known whether there is a phase transition for $N_f = 1$ and if so at what order. For the theory we are studying the relevant chiral symmetry is also $U(1)_B \times U(1)_A$, so it is precisely this $N_f=1$, large $N_c$ case that is closest to the supersymmetric theory we have control over. The same symmetry also governs a single staggered fermion in
the strong coupling limit. Lattice results for that theory \cite{Boyd:1991fb} indicate a second order transition.

\begin{figure}
\center
\begin{tabular}{cc}
\includegraphics[width=0.9\textwidth]{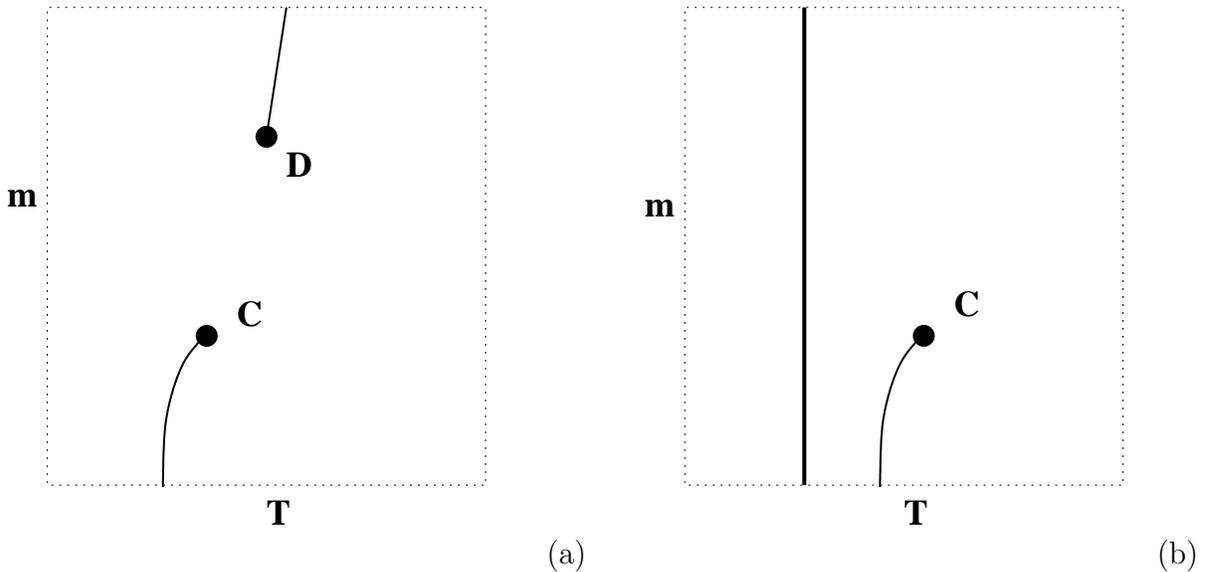} \\
(a) & (b)
\end{tabular}
\caption{\label{qcd}(a.) Phase
Diagram of QCD with $N_c = 3$, $N_f = 2+1$ (b.) $N_c \ra \infty$, $N_f = 2+1$}
\end{figure}

\begin{figure} \center
\begin{tabular}{c}
\includegraphics[width=0.4\textwidth]{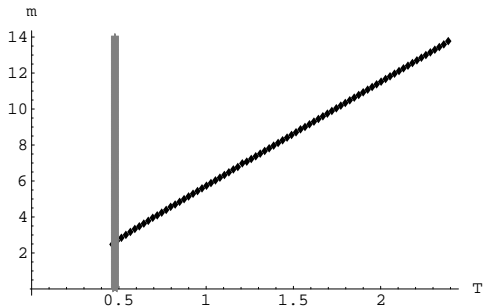}\\
\end{tabular}
\caption{\label{phasediagram}The phase diagram for flavored $\N = 4$
Super-Yang Mills on a sphere as a function of temperature and hyper
mass.}
\end{figure}

Next we wish give the quarks nonzero masses and ask what happens to
the deconfinement and chiral phase transitions as we dial the mass.
Dialing quark masses is not possible in nature but is possible on
the lattice, hence we turn to lattice results although we will only
present them schematically. The phase diagram for $N_C = 3$ and $N_f = 2 + 1$, is shown in fig. \ref{qcd} (a.)
\cite{Pisarski:1994yp}. Here 2+1 means the quark
masses are held at fixed ratios $m = m_u = m_d = r m_s$ for some $r
<1$. What appears are two lines of first-order phase transitions,
each ending in a critical point. The lower line is that of chiral
transitions, ending in the point C; the upper is the line of
deconfinement transitions, ending in the point D.

The chiral transition is identified by a discontinuous jump in the chiral
condensate. Only on the $m=0$ axis, where the chiral symmetry really
is a symmetry, is this a genuine symmetry breaking transition.

The deconfinement transition is identified by a discontinuous jump in the Polyakov
loop, which is the time-ordered exponential of the holonomy of the
gauge field around the time circle. This measures the response of
the system to an infinitely massive source of color charge (a
quark), and may be written

\beq
\<\Omega(x)\> = \< tr P exp( i \int d\t A_0 ) \> \sim exp(- \Delta F / T)
\eeq

where $\Delta F$ is the difference in free energy between two equilibrium states, one with an infinitely massive test quark and one without. For pure glue in the confined phase this difference is infinite and hence $\< \Omega \> = 0$. One can think of this as the statement that the color flux lines from the test quark have no place to end except infinity and hence because the flux lines have finite energy density the total energy is infinite. In the deconfined phase, the flux lines are screened, hence $\< \Omega \>$ is finite. Adding dynamical quarks, massive or otherwise, changes this since now flux lines from the test quark have someplace to end. In other words, with dynamical quarks $\< \Omega \>$ is always nonzero, although it may still jump discontinuously. Only in the $m \ra \infty$ limit where the quarks decouple is the pure glue story recovered.

In summary, both order parameters are nonzero everywhere on the interior of fig. \ref{qcd} (a.). At $m = \infty$ where the quarks decouple, the deconfinement transition occurs at roughly $T_d \sim \Lambda_{QCD} \sim 175 MeV$. Similarly, for $m=0$ where the chiral symmetry is an exact symmetry of the Lagrangian, the chiral transition must occur at the same scale (that then being the only scale in the theory). The horizontal line of realistic quark masses falls between points C and D, hence a smooth crossover rather than a phase transition is expected in the real world.

Now we consider the same diagram but with $N_c \ra \infty$, $N_f$
finite. The answer, again only qualitative, is depicted in fig.
\ref{qcd} (b.). The deconfinement transition, being order $N_{c}^2$,
does not care about the quark masses and is simply a vertical line. We have drawn the chiral transition line the same way although in principle this need not be the case. Large-$N_c$
arguments indicate that the chiral condensate is independent of $T$
in the confining phase \cite{Neri:1983ic,Pisarski:1983db} and hence
the chiral transition must occur at a temperature $T_{ch} \geq T_d$,
we have thus drawn the deconfinement line to the left of the chiral
transition.

In this paper we study the chiral transition in a theory that is in
some sense simpler than QCD: large-$N_c$ $\mathcal{N}=2$ SYM on a
3-sphere. At zero temperature and zero mass, conformal invariance
precludes a nonzero chiral condensate, but finite temperature and
finite mass both break conformal invariance. We will draw the phase
diagram for this theory in the plane of hyper mass versus
temperature, identifying the line of a first-order chiral phase
transition, which occurs only in the deconfined phase, i.e. above
the Hawking-Page temperature. This transition is, in strict
thermodynamic terms, not a chiral symmetry breaking transition, but
the chiral condensate does jump by a finite amount so for simplicity
we will refer to it as a chiral transition. In contrast to QCD we
will see that chiral symmetry is unbroken for zero mass at all
temperatures, so the deconfinement transition does not come with a
standard chiral symmetry breaking transition. Another key difference
from QCD will be an independence on the number $N_f$. Where the
order of the chiral transition in QCD depended on $N_f$, in our
theory it is first order for any finite $N_f$ such that $N_f / N_c
\ra 0$. The ${\cal N}=2$ Lagrangian explicitly breaks the
non-abelian chiral symmetry down to its vector subgroup due to a
Yukawa coupling, so the chiral symmetry breaking we are analyzing is
a breaking of $U(1)_A \times U(1)_B$ down to $U(1)_B$ as in $N_f=1$ QCD.

The thermodynamics of large-$N_c$ $\N = 4$ SYM on a 3-sphere at zero 't
Hooft coupling, but with a Gauss' law constraint requiring color singlet
states, was studied by Sundborg \cite{Sundborg:1999ue}, who found that the
deconfinement transition was first order. Aharony et. al. \cite{Aharony:2003sx} studied the same theory at small but
finite coupling and found that the deconfinement transition could be
either a single first-order transition or two continuous transitions. For pure Yang-Mills on
a 3-sphere, the former scenario was shown to be the right one
\cite{Aharony:2005bq}. Schnitzer \cite{Schnitzer:2004qt} introduced flavor
but with $N_f / N_c$ finite, finding that the deconfinement transition
becomes third order.

The chiral transition of the D3-D7 system for the $\N = 2$ theory in
flat space was studied in \cite{Babington:2003vm,Kirsch:2004km}. A
first-order transition was found \cite{Kirsch:2004km,
Apreda:2005yz}\footnote{While we 
were finishing this manuscript two papers appeared
which also reanalyzed the phase transition in this 
theory, \cite{Mateos:2006nu} and \cite{Albash:2006ew}. In both papers the
picture of a first order phase transition as found 
in \cite{Kirsch:2004km, Apreda:2005yz} is 
confirmed. Like us, they simplify the
numerics by integrating outward instead of shooting inward. In addition \cite{Mateos:2006nu} argues that when the flavor brane touches the horizon the probe action acquires a new scaling symmetry and a self-similar structure. Their general analysis of Dp/Dq systems corresponding to large-$N_c$, (p+1)-dimensional thermal
gauge theory in infinite volume with fundamental matter confined to
a (q+1)-dimensional defect also shows that such phase transitions
must be first order. Our results are an extension of this since we
consider finite volume and find again a first-order transition in the deconfined phase.}.
A similar phase transition was also observed for flavor branes
in the supergravity dual of confining gauge theories
\cite{Kruczenski:2003uq}.
The phase diagram must by conformality be linear: the quark mass is
the only scale that can set a transition temperature. As we will
argue below, this infinite volume result can be interpreted as a
high-temperature result for the theory on the 3-sphere, and indeed
we will find a straight line at high temperatures, providing a useful check of our methods.

Our final result for the chiral transition is the phase diagram in
figure (\ref{phasediagram}). The vertical line is the Hawking-Page deconfinement
temperature, $T_{HP} = 3/2\p$ (in units where the AdS radius is
one), which comes from order $N_c^2$ dynamics and is independent of the quark mass.
Below $T_{HP}$, the flavor brane action is independent of
temperature, hence whatever occurs at $T=0$ can be extended up to
$T_{HP}$. This is consistent with the large-$N_c$ field theory arguments that the chiral condensate is independent of temperature in the confining phase
\cite{Neri:1983ic, Pisarski:1983db}. We find no first-order phase transition as a function of temperature below $T_{HP}$.  This comes as a surprise since at zero temperature we see a topology change for the D7 brane as a function of mass.

Above $T_{HP}$, the flavor brane action depends on the temperature via the
presence of the black hole horizon. In this temperature regime, the chiral condensate and free energy exhibit behavior characteristic of a first order transition as $m$ increases, i.e. the first derivative of the free energy is
discontinuous and so on. We have drawn the first-order chiral transition line. Below $T_{HP}$, the chiral condensate smoothly changes, as does the free energy, as $m \ra \infty$, so we draw no first-order transition line. Everywhere on the $m=0$ axis, we find numerically that the chiral condensate is zero.

This paper is organized as follows. In section \ref{flavor} we
review the inclusion of flavor branes into the ${\cal N}=4$ theory
and identify the chiral symmetry in this theory and its order
parameter. In section \ref{action} we write down the probe brane
action and counterterms, explain how we extract the condensate and
free energy using holographic renormalization, and explain the
boundary conditions for our numerics. In section \ref{flat} we reproduce known results in infinite volume. In section
\ref{curved} we present numerical results for the embeddings,
condensate, free energy and finally for the phase diagram in the finite volume case. We
conclude in section \ref{conclusion} with possibilities for extensions of this study.

%%%%%%%%%%%%%%%%%%%%%%%%%%%%%%%%%%%%%%%%%%%%%%%%
\section{Adding Flavor to AdS/CFT}\label{flavor}
%%%%%%%%%%%%%%%%%%%%%%%%%%%%%%%%%%%%%%%%%%%%%%%%

We will now describe our system in greater detail, clarifying both
the D-brane construction and the symmetries of the boundary theory.
The AdS/CFT construction starts with a stack of coincident D3-branes
in flat ten-dimensional space in the limit where the number of
D3-branes, $N_c$, goes to infinity and the spacetime in the
near-horizon limit becomes $AdS_5 \times S^5$. The limit of zero
string length $\alpha' \rightarrow 0$ then decouples open and closed
strings, resulting in the now-standard correspondence between the
low-energy effective theory of closed strings, ten-dimensional
supergravity (SUGRA) in this background, and the low-energy
effective theory of open string modes on the D3 worldvolume,
$\mathcal{N} = 4$ super-Yang-Mills (SYM) theory in four dimensions
in the large-$N_c$, strong 't Hooft coupling regime. This
four-dimensional conformal field theory (CFT) will ``live" in a
spacetime with the topology of the $AdS_5$ boundary. Our interest
here is in global thermal $AdS$ (and $AdS$-Schwarzschild)
coordinates, with boundary topology $S^1 \times S^3$, which we will
call the ``curved case", in distinction to the infinite-volume ``flat
case", that is, Poincar\'{e} patch coordinates with boundary topology
$S^1 \times R^3$. We will use the limit of infinite volume to
check our methods against known flat case answers.

Thermal AdS is believed to undergo a first-order phase transition
\cite{Hawking:1982dh,Witten:1998zw} called the Hawking-Page
transition. At low temperatures the picture is of thermal radiation
in equilibrium with itself (assuming perfectly reflecting boundary
conditions at the AdS boundary), whereas when the temperature rises
the spacetime undergoes a topology change to the AdS-Schwarzschild
solution, or in other words a black hole condenses. This transition
is only apparent in global AdS coordinates, while the flat case is
strictly in the high-temperature black hole phase.

In the dual $\mathcal{N} = 4$ SYM boundary theory, the Hawking-Page
transition corresponds to a large-$N_c$ deconfinement transition.
``Deconfinement" here means the spontaneous breaking of the center symmetry
and the associated behavior of that symmetry's order parameter, the
Polyakov loop, which jumps discontinuously (from zero to nonzero) as the
temperature rises. This transition is only apparent for the theory on a
compact space, where the inverse $S^3$ radius can set a transition temperature.

In fact, the ratio of the $S^1$ and $S^3$ radii is the only meaningful
number so long as the theory is conformal. Letting $R_1$ be the $S^1$
radius (inversely proportional to the temperature) and $R_3$ be the $S^3$
radius, the infinite-volume limit $R_1 / R_3 \ra 0$ is equivalent to the
infinite-temperature limit in which $R_1 \ra 0$. In this sense we can compare our finite-volume results, at high-\textit{temperature}, with known infinite-\textit{volume}, finite-temperature results.

Introducing D7-branes orthogonal to the initial D3's in four directions
introduces new open string degrees of freedom in the D3 worldvolume
theory, from D3-D7 strings, and breaks half the supersymmetry. We will
denote the number of D7's as $N_f$. The so-called probe limit consists of
keeping $N_f$ fixed as $N_c \rightarrow \infty$. In this limit, the
backreaction of the D7's on the geometry is negligible and hence the
result in the near-horizon limit is $N_f$ probe D7-branes embedded in
$AdS_5 \times S^5$. From the field theory perspective, knowing that the
$\mathcal{N} = 4$ supersymmetry must be broken to $\mathcal{N} = 2$ and
from Chan-Paton factors of D3-D7 strings, one can argue that this
corresponds to adding matter in the fundamental representation of the
gauge group to the $\mathcal{N} = 4$ theory. Specifically, a number $N_f$ of
flavors in the form of $\mathcal{N} = 2$ hypers have been added.

From the field theory point of view, these hypers may be massive without
breaking supersymmetry although of course this will break conformal
invariance. This is manifest in the dual brane picture.
In this setup, the D7 wraps all of $AdS_5$ as well as a trivial cycle inside the $S^5$, namely an $S^3$. A stable
``slipping mode" then exists, which is simply a scalar field on the
worldvolume of the D7 that depends only on the $AdS_5$ radial coordinate \cite{Karch:2002sh}.
The $S^3$ that the D7 wraps inside the $S^5$ may ``slip off" the $S^5$, as
allowed by topology, i.e. contract to a point. If this occurs at a finite
value of the radial coordinate, the hyper in the field theory can be shown
to have a nonzero mass. The zero mass conformal case is then simply a D7
wrapping the same equatorial $S^3$ for all values of the radial
coordinate.

Indeed, via the AdS/CFT dictionary \cite{Karch:2002sh}, this scalar
slipping mode encodes both the mass of the hyper and the chiral condensate
as a function of the mass, as we show below. The key observation is that
the D7 worldvolume scalar has a mass (as a scalar in $AdS_5$) of $m^2 =
-3$ (negative but above the Breitenlohner-Freedman bound \cite{Breitenlohner:1982jf}) and hence must be dual to a CFT
operator of dimension 3 or 1, built from hyper fields. The only such
unitary operator is a fermion bilinear. AdS/CFT then states that the
source for the dual operator (the fermion mass) and the value of the chiral condensate can be
extracted from the asymptotic expansion of the scalar. These statements are made precise in section \ref{holorg}.

Our goal is thus to map the phase diagram of this $\mathcal{N} = 2$,
non-conformal theory in the plane of mass versus temperature using this
chiral condensate as our order parameter. In what sense is this chiral
condensate an order parameter, however? That is, what are the symmetries
of our theory and what symmetry breaking does this order parameter detect?

$\mathcal{N} = 4$ SYM has an $SU(4) \simeq SO(6)$ R-symmetry. The
hypers break the R-symmetry to $SO(4) \times SO(2)$. This is easy to
see on the gravity side: the D7 wraps an $S^3$ inside the $S^5$,
breaking the $SO(6)$ isometry to the $SO(4)$ of the $S^3$ and an
$SO(2)$. This $SO(2) \simeq U(1)$ is a chiral symmetry in that the
hyper fermions, which have opposite chirality, carry opposite
charges under it. With $N_f$ flavors, the theory has in total a
global $U(N_f) \times SO(4)_R \times SO(2)_R 
\simeq SU(N_f) \times U(1)_B \times SO(4)_R \times U(1)_A$. The $U(1)_A$ is
our chiral symmetry in this case. At finite $N_c$, this $U(1)_A$
would be anomalous but as in large-$N_C$ QCD the axial anomaly is
suppressed as $N_c \ra \infty$. Without the Yukawa
interactions required by $\cn=2$ supersymmetry the $U(N_f)$ global
symmetry would be enhanced to a $U(N_f)_L \times U(N_f)_R$.

In this paper, we consider only $N_f = 1$. This poses no problem on the gravity side: for this system, at zero
mass, the factor of $N_f$ would simply appear in front of the D7 action
and would not affect the equations of 
motion\footnote{One subtelty that would however be
present for $N_f \geq 2$ is that the configuration of lowest
free energy might be found on the Higgs branch \cite{Apreda:2005yz}.
For $N_f=1$ there is no Higgs branch since all hyper scalars
have a quartic potential.}. These statements remain true at
nonzero mass so long as all the flavor branes end at the same radius. A
construction with different D7's ending at different radial values is
possible, where the details of the spectrum would fix what remains of the
$U(N_f)$, but we will not consider such a situation here (though it would
be a straightforward extension).

%%%%%%%%%%%%%%%%%%%%%%%%%%%%%%%%%%%%%%%%%%%%%%
\section{The Probe Brane Action}\label{action}
%%%%%%%%%%%%%%%%%%%%%%%%%%%%%%%%%%%%%%%%%%%%%%

%%%%%%%%%%%%%%%%%%%%%%%%%%%%%%%%%%%%%%%%%%%%%%%%%%%%%%%%%%%%%%%%%%%%%%%
\subsection{Fefferman-Graham Coordinates}\label{fgsec}
%%%%%%%%%%%%%%%%%%%%%%%%%%%%%%%%%%%%%%%%%%%%%%%%%%%%%%%%%%%%%%%%%%%%%%%

We will now write the action for the D7 slipping mode and show how to extract the mass and chiral
condensate from solutions of the equation of motion (EOM). We also describe the boundary
conditions in detail as these are somewhat different from those in
\cite{Babington:2003vm,Kirsch:2004km}. In these references and in
much of the literature, a shooting method is employed, which is
appropriate given only UV, or boundary, data. We impose boundary
conditions in the IR, which is much simpler for numerics. In section \ref{flat} we will show that our IR boundary conditions reproduce the results of \cite{Babington:2003vm,Kirsch:2004km} for the flat case. Boundary conditions similar to ours were used recently in \cite{Albash:2006ew} and also reproduced these results.

In units where we set the curvature radius $L$ to one
the $AdS_5$ metric in global coordinates has the form
\beq
ds^2 = G_{AB} dx^A dx^B = \frac{dr^2}{f(r)} + f(r) d\tau^2 + r^2 d\Omega_{3}^2
\eeq

\bea f(r)  &=& \left \{
\begin{array}{cccl}
        && 1 + r^2
 & \, \, \, \,\mbox{thermal AdS} \\

        && 1 + r^2 - \frac{M^2}{r^2}
 & \, \, \, \,\mbox{AdS-Schwarzschild}
\end{array}
\right . \eea
$d\Omega_{p}^2$ is the standard metric for a $p$-sphere. Here the
time coordinate $\tau$ has Euclidean signature and is periodic with
period $2 \pi T$ for temperature $T$. The parameter $M$ is related
to the black hole mass $m_{bh}$ as $M=\frac{8 G_N}{3 \pi} m_{bh}$ and to the temperature as $M = \p T$.
For AdS-Schwarzschild, the horizon is at $r_+$, the largest solution
to the equation $f(r_+)=0$. The flat case is the same with no 1 in
$f(r)$ and with Euclidean 3-space instead of the $S^3$.

In the above coordinates, the boundary is at $r \ra \infty$. This makes
extracting asymptotics from numerics difficult. These coordinates are also
not well-suited to holographic renormalization. We thus switch to a
Fefferman-Graham \cite{Fefferman} coordinate system, where the metric takes the
form

\beq
ds^2 = \frac{dz^2}{z^2} + \frac{1}{z^2} g_{ij}(z,\vec{x},M) dx^i dx^j
\eeq

\beq
g_{ij}(z,\vec{x},M) dx^i dx^j = \frac{1}{4} (1- z^4(1+4 M^4))^2 F(z,M)^{-1} d\t^2 + \frac{1}{4} F(z,M) d\O^{2}_3
\eeq

\beq
F(z,M) = 1- 2 z^2 + z^4(1+4 M^4)
\eeq

\noindent with radial coordinate $z$ and $S^1 \times S^3$ coordinates
$x^i$. The boundary is now at $z=0$ and the horizon is at $z_+ = (1+4
M^4)^{-1/4}$. Taking $M=0$ gives $g_{ij}$ for thermal AdS with center at $z=1$. Notice also that in our conventions our boundary $S^3$ has radius $1/2$.

The metric on the slice has an asymptotic expansion

\bea
g_{ij}(z,\vec{x},M) & = & g^{0}_{ij} + g^{2}_{ij} z^2 + g^{4}_{ij} z^4 + O(z^6) \\
g^{0}_{ij} & = & \frac{1}{4} diag(1,S^3) \\
g^{2}_{ij} & = & -\frac{1}{2} diag(-1,S^3) \\
g^{4}_{ij} & = & \frac{1}{4} diag(1 - 12 M^4, (1 + 4 M^4) S^3)
\eea

where ``$S^3$" denotes the $S^3$ metric components. The $S^5$ metric is

\beq
d\Omega_{5}^2 = d\theta^2 + \sin^{2}\theta d\psi^2 + \cos^{2}\theta d\Omega_{3}^2
\eeq

The D7 then wraps the whole of $AdS_5$ and the $S^3$ inside the $S^5$. The
D7 slipping mode is the worldvolume scalar field $\theta = \theta(z)$,
which is a function only of the radial coordinate to preserve the Lorentz
invariance of the boundary theory. If $\theta(z_0) = \frac{\pi}{2}$ for
some finite $z_0$, then the $S^3$ has ``slipped off" the $S^5$.

The D7 action is the Dirac-Born-Infeld (DBI) action, which in this
background is simply the volume of the probe brane:

\beq
S_{D7} = T_{D7} \int d^8 \xi \sqrt{det P[G]}
\eeq

where $P[G]$ is the pullback of the spacetime metric. With the
$\th(z)$ ansatz, the relevant part of the action is

\beq
S_{D7} = \int dz \frac{1}{16} \frac{1}{z^5} F(z,M) (1 - z^4(1+4 M^4)) \cos^{3}\th(z) \sqrt{1 + z^2 \th'(z)^2}
\eeq

\noindent Again thermal AdS is the same with $M=0$. Our objective is to solve the $\th(z)$ EOM. We show in section \ref{holorg} how to extract the mass and chiral condensate from a solution and in section \ref{bc} we explain the boundary conditions we impose on solutions.

%%%%%%%%%%%%%%%%%%%%%%%%%%%%%%%%%%%%%%%%%%%%%%%%%%%%%%%%%%%%%%%%%%%%%%%
\subsection{Operator normalization}\label{normalization}
%%%%%%%%%%%%%%%%%%%%%%%%%%%%%%%%%%%%%%%%%%%%%%%%%%%%%%%%%%%%%%%%%%%%%%%

While we have already identified the operator dual to the slipping
mode $\theta$ as the dimension 3 fermion bilinear, there
is still some freedom in the overall normalization. The correct normalization of the operator can be fixed by studying the 2-point function at zero temperature. Performing the
integral over the $S^3$ in the 7-brane action gives a 5d action with
prefactor

\beq
\frac{1}{g_5^2} = T_{D7} V_{S^3} = \frac{N_f}{g_s 2 \pi^7 (\alpha')^4} (2 \pi^2) = \frac{\lambda N_c N_f}{(2 \pi)^4}.
\eeq

\noindent where we used that in our $L=1$ units $(\alpha')^{-2} = \lambda
= g^2_{YM} N_c = 4 \pi g_s N_c$. According to the AdS/CFT dictionary
the dual operator $O$ has a 2-pt function

\beq
\langle O(x) O(0)
\rangle = \frac{1}{g_5^2} \frac{2 \Delta -d}{\pi^{d/2}}
\frac{\Gamma(\Delta)}{\Gamma(\Delta-d/2)} \frac{1}{x^{2\D}} =
\frac{\lambda N_c N_f}{4 \pi^6} \frac{1}{x^6}.
\eeq

In the field theory we are interested in the source term (the mass) and a vacuum
expectation value for the operator $\tilde{O} = \bar{\lambda}
\lambda$ where $\lambda$ is a canonically normalized fermion. So far
we have only established that $O = C \tilde{O}$ for some
normalization constant $C$, but in order to interpret our results in
the field theory we need to know $C$. One way to fix $C$ is to
compare the two point functions of $O$ and $\tilde{O}$. Since
$\tilde{O}$ is a protected operator (which follows from the non-renormalization of the flavor current 2-pt function established in \cite{Aharony:1999rz} by supersymmetry), its 2-pt function can be evaluated in the free theory. For a canonically-normalized fermion
the position space propagator is simply $\langle \bar{\lambda}(x)
\lambda(0) \rangle = \frac{1}{2 \pi^2} \frac{\slashed{x}}{x^4}$
times flavor and color delta functions. So in the free theory we
then simply get from Wick contractions

\beq
\< \tilde{O}(x)
\tilde{O}(0) \> = \frac{N_f N_c}{(2 \pi^2)^2} \frac{\Tr( \slashed{x}
\slashed{x})}{x^8} = \frac{N_f N_c}{\pi^4} \frac{1}{x^6}
\eeq

From which we can conclude that

\beq
O = \frac{\sqrt{\lambda}}{2\pi} \tilde{O}.
\eeq

From this relation it becomes clear that all masses and condensates
we find in this paper have to be multiplied by
$\frac{\sqrt{\lambda}}{2\pi}$ to translate into field theory units.
The phase transitions we study will happen at masses of order
$\sqrt{\lambda} T$. This is the natural scale since the mesons in
the theory only have masses of order $\frac{m}{\sqrt{\lambda}}$
\cite{Kruczenski:2003be}. At zero temperature the mass we obtain
this way also agrees with the energy of a straight string stretching
from the horizon to the flavor brane.

Note that the scaling of $\frac{1}{g_5^2}$ with $N_f$ and $N_c$ is
expected simply from large $N_c$ counting rules. The factor of $\lambda$
however is a little surprising at first sight. Besides its effect
on the normalization of the operators it will for example lead to
free energies proportional to $\lambda$ at large $\lambda$, possibly indicating
that in addition to the free value there is a 1-loop quantum correction
but no further contributions from higher loops.

%%%%%%%%%%%%%%%%%%%%%%%%%%%%%%%%%%%%%%%%%%%%%%%%%%%%%%%%%%%%%%%%%%%%%%%
\subsection{Holographic Renormalization and Free Energy}\label{holorg}
%%%%%%%%%%%%%%%%%%%%%%%%%%%%%%%%%%%%%%%%%%%%%%%%%%%%%%%%%%%%%%%%%%%%%%%

Given a solution of the EOM we need to extract the hyper mass and
the chiral condensate. We use the method of holographic
renormalization (which we call ``holo-rg") to do this as well as
compute the on-shell action. We will now briefly review the holo-rg
procedure
\cite{Henningson:1998gx,Henningson:1998ey,Balasubramanian:1999re,deHaro:2000xn}.

The precise statement of AdS/CFT equates the 10D on-shell SUGRA action
with the generating functional of the boundary field theory. The leading
asymptotic value of a bulk field serves as a source for the corresponding
field theory operator, i.e. derivatives of the SUGRA action w.r.t. leading
asymptotic values will give boundary correlators. This will be made
explicit for our $\th(z)$ field below.

Both the SUGRA action and the field theory generating functional are
formally infinite: the SUGRA action suffers IR divergences while the field
theory has UV divergences. Holo-rg proceeds first by regulating the SUGRA
action. This means integrating not to the $z=0$ boundary but only to
$z=\e$. Local covariant counterterms are then added on the $z=\e$ slice,
built from the metric induced on the slice, $\g_{ij}$, curvature
invariants thereof and fields on the slice, for instance our $\th(\e)$.
The coefficients of these counterterms are fixed by requiring all
divergences to cancel. Functional derivatives w.r.t. asymptotic values can
then be taken, followed by removal of the regulator, $\e \ra 0$, yielding
renormalized boundary correlators. Covariance and all other symmetries can
be maintained throughout this procedure. Holo-rg also gives us a way to compute a renormalized free energy of our $\N=2$ theory, since via AdS/CFT this is just equivalent to the on-shell SUGRA action. We find this much more elegant, in principle and in practice, than a background subtraction technique.

Now to make all of this explicit for our system. The holo-rg
procedure for precisely this system, a probe D7 in $AdS_5 \times
S^5$, was worked out in \cite{Karch:2005ms}, so we quote those results. According to the AdS/CFT dictionary, the mass
and condensate for the hyper fermions come from the asymptotic expansion

\beq\label{sol}
\th(z) = \th_0 z + \th_2 z^3 + \TH_2 z^3 log z + \ldots
\eeq

where the mass will be given by $\th_0$ and the condensate will be fixed by $\th_0$ and $\th_2$. In this case, the log coefficient $\TH_2 = \frac{1}{12} R_0 \th_0$ \cite{Karch:2005ms} with $R_0$ the Ricci scalar of $g^0$. For us $R_0 = 24$ so $\TH_2 = 2 \th_0$. The counterterms for this system are given in \cite{Karch:2005ms}\footnote{This reference used a convention that AdS space had constant \textit{positive} curvature.  We use the opposite convention, hence our counterterms have $R_{\g} \ra - R_{\g}$ relative to those in \cite{Karch:2005ms}.}:

\bea
L_1 &=& - \frac{1}{4} \sqrt{\g} \\
L_2 &=& \frac{1}{48} \sqrt{\g} R_{\g} \\
L_3 &=& - \log(\e) \sqrt{\g} \frac{1}{32} ( R_{ij} R^{ij} - \frac{1}{3} R_{\g}^2 ) \\
L_4 &=& \frac{1}{2} \sqrt{\g} \th^2(\e) \\
L_5 &=& \frac{1}{12} \log\th(\e) \sqrt{\g} R_{\g} \th(\e)^2 \\
L_{f} & = & -\frac{5}{12} \sqrt{\g} \th(\e)^4
\eea

We will denote the subtracted action as $S_{sub} = S_{D7} + \sum_i L_i$ and the renormalized action as $S_{ren} = \lim_{\e \ra 0} S_{sub}$. Notice that $\sqrt{\g} = \e^{-4} \sqrt{g}$ and $R_{\g} = \e^2 R_{g}$ and to leading order $\th(\e) \sim \th_0 \e$, so the last counterterm is finite. The first three counterterms are needed to renormalize the volume of $AdS_5$. The divergent piece of $L_3$ has coefficient $R^{0}_{ij} R_{0}^{ij} - \frac{1}{3} R_{0}^2$. For our $g^{0}_{ij}$, this quantity vanishes and hence $L_3$ is not needed. $L_4$ and $L_5$ are the counterterms for a free scalar in $AdS_5$. The
coefficient of $L_{f}$ is fixed by requiring a supersymmetric renormalization scheme \cite{Karch:2005ms}. Our $L_5$ is actually slightly different from the one in \cite{Karch:2005ms}, which has $\log(\e)$ instead of $\log\th(\e)$. This is so a large-mass expansion correctly reproduces the flat case answer, as we now show. The condensate is given by \cite{Karch:2005ms}

\bea
\< \bar{q} q \> & = & \lim_{\e \ra 0} \frac{1}{\e^3 \sqrt{\g}} \frac{\d S_{sub}}{\d \th(\e)} \\ & = & -2 \th_2 + \frac{1}{3} \th_{0}^3 + \frac{1}{12} R_0 \th_{0} \log(\th_{0}^2).
\eea

Let us focus on zero temperature. At large mass any finite volume effects should be negligible and we should recover the flat space results, that is the leading piece in both vev and free energy vanish \cite{Karch:2005ms}. Writing the vev as a power series

\beq
\langle \bar{q} q \rangle =  m^3 \left ( o_0 + o_1 \frac{1}{(m R_3)^2} + o_2 \frac{1}{(m R_3)^4} + o_3 \frac{1}{(m R_3)^6} + \ldots \right )
\eeq

\noindent the coefficients $o_0$, $o_1$, $\ldots $ can be solved for order by order in $\frac{1}{m R_3}$. It is convenient to reinstate $R_3$ in the bulk metric:

\beq
ds^2 = \frac{dz^2}{z^2} - \frac{1}{4} \left ( \frac{1}{z^2} + \frac{2}{R_3^2} + \frac{z^2}{R_3^4} \right ) R_3^2 dt^2 + \frac{1}{4} \left (\frac{1}{z^2} - \frac{2}{R_3^2} + \frac{z^2}{R_3^4} \right ) R_3^2 d \Omega_3^2.
\eeq

The equations of motion following from the DBI action

\beq S_{D7} = R_3^4 \int \frac{d^5x}{z^5} \frac{1}{16} \cos^3 \theta(z) \, \sqrt{1 + \frac{2z^2}{R_3^2} + \frac{z^4}{R_3^4} } \,  \left ( 1 - \frac{2 z^2}{R_3^2} + \frac{z^4}{R_3^4} \right )^{3/2} \sqrt{1 + z^2 \theta'(z)^2}
\eeq

can be solved order by order in powers of $\frac{1}{R_3}$,

\beq
\sin(\theta(z)) = arg_0(z) + \frac{1}{R_3^2} arg_1(z) + \ldots.
\eeq

To leading order one reobtains the flat case solution $arg_0(z) =  c z$ and hence $o_0=0$ since the vev in the flat case solution was zero.
 
Expanding to order $R_{3}^{-2}$ one obtains

\beq
arg_1(z) = \frac{1}{c} \frac{ z - c^2 z^3 + c^2 z^3 \log (z c)^2 }{ 1- c^2 z^2 }.
\eeq

First note that the prefactor of $\frac{1}{c}$ compared to $c$ in the 0-th order solution indeed nicely combines with the $\frac{1}{R_3^2}$ to form the dimensionless expansion parameter $\frac{1}{c^2 R_3^2}$. Also note that the solution contains a $\log c$ term, which nicely combines with the $\log z$ into $\log z c$. It is straightforward to calculate $\langle \bar{q} q \rangle$ for this analytic solution and we find with the counterterms above that $o_1=0$, so that in the large mass limit the vev vanishes\footnote{Note that $o_2$ and all higher terms multiply a negative power of $m$, so that those contributions to the vev vanish at large mass even for non-zero coefficients. Indeed we have also worked out the 2nd order correction to $\theta(z)$ and found a non-zero $o_2$.}. It is this analytic answer for the first order correction that demands the particular form of $L_5$ we use. (Note that this only works because the log terms combined in this particular way.) Any additional finite counterterm such as $\sqrt{\gamma} R_{\g} \theta^2$ would spoil the vanishing of the $o_2$ term so all the counterterms are specified uniquely by the requirement that the vev vanishes in the large mass limit. It is easy to see that the analysis also carries through at nonzero temperature, where the metric only gets modified at order $\frac{1}{R_3^4}$ so that finite temperature does not affect $o_0$ and $o_1$.

With these counterterms implemented in the numerics the condensate at very large masses doesn't go exactly to zero and even dips negative. Our analytic solution makes it clear that this is a numerical artifact. It is easy to understand that the numerics becomes questionable at large mass, since in that case the position of the flavor brane pushes into the region in which we try to fit the numerical solution against the expected near boundary behavior in asymptotically AdS space.

Even at small mass, however, we see the condensate becoming negative (at least for the curved case: see figures (\ref{vevtads}) and (\ref{vevadsbh})). While at first sight this may seem disturbing, it is not forbidden. In non-relativistic quantum mechanics first-order perturbation theory always overestimates the free energy (ground state energy), \textit{i.e.} the first-order correction is always positive. The second order correction is always negative and therefore the free energy is always a concave increasing function. A relativistic field theory, however, requires renormalization. Counterterms proportional to $R_0 m^2$ and $m^4$ will be present, and depending on the choice of their coefficients (choice of scheme) the corrections to the free energy may make the free energy decrease (and possibly convex). The condensate is the derivative of the free energy and hence may become negative. The scheme dependence of the field theory shows up holographically precisely in $L_5$ and $L_f$. We believe we have fixed these counterterms in a physically well-motivated way and that negative condensates are simply a consequence of this choice of scheme and no cause for alarm.

The free energy of the field theory is given by the on-shell value
of the renormalized action. The order $N_c^2$ contribution to the
free energy comes from the 10d SUGRA action and is untouched by the
introduction of the probe brane. Naively there are two sources of
order $N_f N_c$ corrections, the renormalized value of the on-shell
DBI action as well as the order $\frac{N_f}{N_c}$ correction to the
order $N_c^2$ on-shell bulk SUGRA action due to the backreaction.
The latter however is proportional to the variation of the supergravity
action with respect to the unperturbed background metric and hence
vanishes. The backreaction only enters at order $N_{c}^0$ and the only
order $N_f N_c$ contribution to the free energy comes from the
renormalized on-shell DBI action.

%%%%%%%%%%%%%%%%%%%%%%%%%%%%%%%%%%%%%%%%%%
\subsection{Boundary Conditions}\label{bc}
%%%%%%%%%%%%%%%%%%%%%%%%%%%%%%%%%%%%%%%%%%

We will discuss first the $M=0$ thermal AdS case, with no black hole
horizon. We consider two kinds of solutions. The first kind are D7's
ending at some finite $z_0$, i.e. $\th(z_0) = \pi / 2$. This is our first
boundary condition, and we want a second condition in the IR. We find one
by requiring that the D7 ends smoothly, with no conical deficit. The
pulled-back D7 metric contains terms

\beq
ds^{2}_{D7} \supset (\frac{1}{z^2} + \th'(z)^2 ) dz^2 + \cos^{2} \th(z) d\O_{3}^2
\eeq

\noindent Let $\a = \cos \th(z)$ so that $d\a = - \sin \th(z) \th'(z)$ and hence

\beq
ds^{2}_{D7} \supset \frac{(\frac{1}{z^2} + \th'(z)^2)}{\th'(z)^2 \sin^{2}\th} d\a^2 + \a^2 d\O_{3}^2
\eeq

No conical deficit means the coefficient of $d\a^2$ must be one when $z=z_0$. This gives $z_{0}^{-2} \th '(z_0)^{-2} = 0$ and hence $\th'(z_0) = \infty$. We implement this numerically with a number on the order of $10^3$.

The second kind of solution is for D7's extending to the center of
AdS at $z_0 = 1$. In this case, we choose a boundary condition
$\th(z=1) = \b$ for some angle $0 \leq \b \leq \p / 2$. $\b = 0$ is
the massless case and dialing $\b$ will interpolate smoothly to a D7
ending precisely at the center. The second boundary condition is now
the Neumann condition $\th'(z=1) = 0$.

In AdS-Schwarzschild we also have two kinds of solutions, D7's ending outside the horizon or ending on the horizon. For D7's ending outside the horizon we use the same boundary conditions as those in thermal AdS. For D7's ending on the horizon we use
boundary conditions $\th(z_+) = \b$ and $\th'(z_+) = 0$. The flat case, where the horizon is present for any finite temperature, also has these two types of solutions with these boundary conditions.

In summary, our procedure is this: solve the EOM for $\th(z)$ with these boundary conditions. With this solution, we
can immediately find the mass and condensate for that value of $z_0$ or $\b$. We also plug this solution into $S_{ren}$ to find the free energy. We can then go to the next value of $z_0$ or $\b$, which amounts to dialing the mass. In
this fashion, from the behavior of the condensate and the free energy, we
can identify the chiral phase transition and determine its order.

In much of the literature (for example \cite{Babington:2003vm,
Kirsch:2004km,Evans:2004ia}), IR boundary information such as this
is not used. Instead, a shooting method is used. To justify and
cross-check our method, in the next section we will reproduce some
known results for the flat case using our boundary conditions and
holo-rg. We will also compare our curved case results in the
high-temperature limit to these results as a useful check.

%%%%%%%%%%%%%%%%%%%%%%%%%%%%%%%%%%%%%%
\section{The Flat Case}\label{flat}
%%%%%%%%%%%%%%%%%%%%%%%%%%%%%%%%%%%%%%

The thermodynamics of this $\N = 2$ theory has been studied in \cite{Babington:2003vm,Kirsch:2004km}. We will compute the chiral condensate as a function of the mass, the free energy as a function of the mass and then draw the phase diagram. As suggested in \cite{Babington:2003vm,Kirsch:2004km}, a first-order phase transition does occur at a critical value of the mass. The critical value is $m_{crit} \approx 1.30$ (for $M=1$)\footnote{Our $m_{crit}$ is larger than the $m_{crit} \approx 0.92$ in \cite{Babington:2003vm,Kirsch:2004km,Albash:2006ew} by a factor of $\sqrt{2}$. This comes from our choice of coordinates. The coordinate $r$ in \cite{Albash:2006ew} is related to our $z$ by $z = \frac{1}{\sqrt{2}} \frac{1}{r}$. We verified that this change of coordinates reproduces fig. 2(b) of \cite{Albash:2006ew}. While a change of coordinates clearly cannot change a physical quantity like the mass, different coordinate systems suggest different natural defining functions to extract the boundary metric, $z^2$ in our case but $r^2$ for \cite{Albash:2006ew}. The corresponding boundary metrics differ by a factor of $\sqrt{2}$ for the radius of the $S^3$ and hence the mass measured in units of the inverse radius of the sphere differs by a factor of $\sqrt{2}$ as well.}. The form of the free energy shows this clearly. In this case, our $AdS_5$ metric is

\beq
ds^2 = \frac{dz^2}{z^2} + \frac{1}{z^2} \frac{1}{2} (1 - z^4 M^4)^2 F(z,M)^{-1} d\t^2 + \frac{1}{z^2} \frac{1}{2} F(z,M) d\vec{x}^2
\eeq

\beq
F(z,M) = 1 + z^4 M^4
\eeq

\noindent and the horizon is at $z_+ = 1/M$. An important fact we will need later is that the last term in the metric has a $1/2$ whereas the curved case metric had a $1/4$, hence $z$ in the flat case is $\sim \sqrt{2} z$ in the curved case and the mass we extract from eq. \ref{sol} in the flat case will thus be $1/\sqrt{2}$ times the mass we compute in the curved case. This is important when comparing the phase diagrams in the two cases, in particular their slope, as we do in section \ref{curved}.

The D7 action is now

\beq
S_{D7} = \int dz \frac{1}{4} \frac{1}{z^5} (1 - z^8 M^8) \cos^3\th(z) \sqrt{1 + z^2 \th'(z)^2} .
\eeq

The counterterms are given above but with $L_2 = L_3 = L_5 = 0$ because $R_{\g} = 0$.

\begin{figure}
\center
\begin{tabular}{cc}
\includegraphics[width=0.4\textwidth]{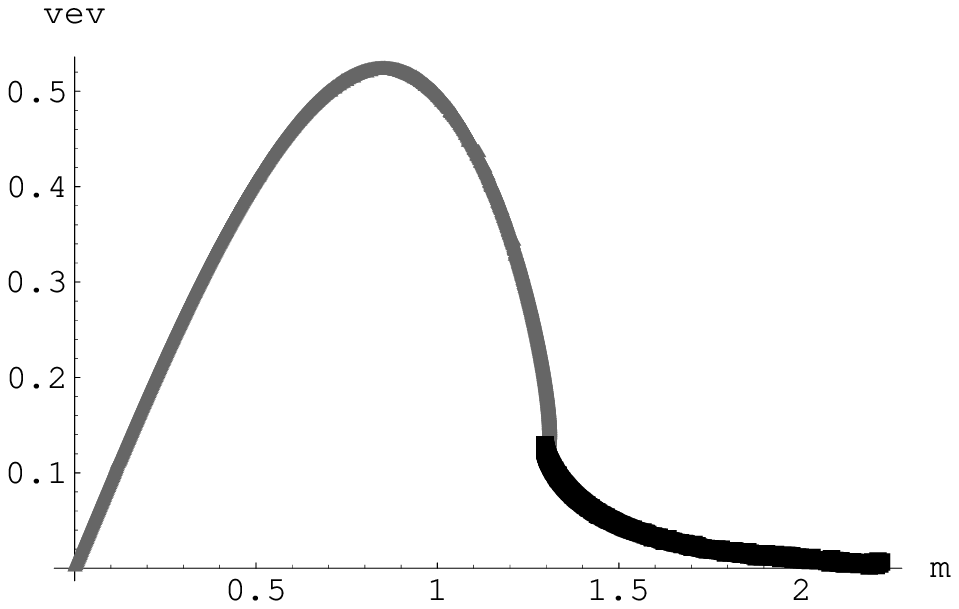} &
\includegraphics[width=0.4\textwidth]{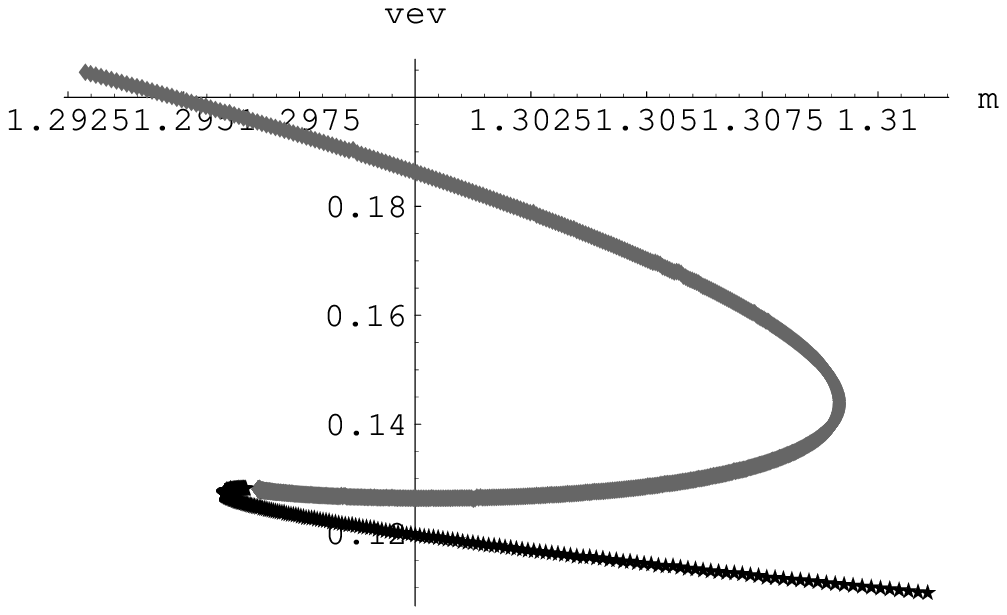}\\
(a) & (b)
\end{tabular}
\caption{\label{flatv}(a.) Chiral Condensate as a function of mass ($M=1$);
gray and black curves indicate solutions ending inside and outside the
horizon respectively (b.) Close-up view of (a.)}
\end{figure}

\begin{figure}
\center
\begin{tabular}{cc}
\includegraphics[width=0.4\textwidth]{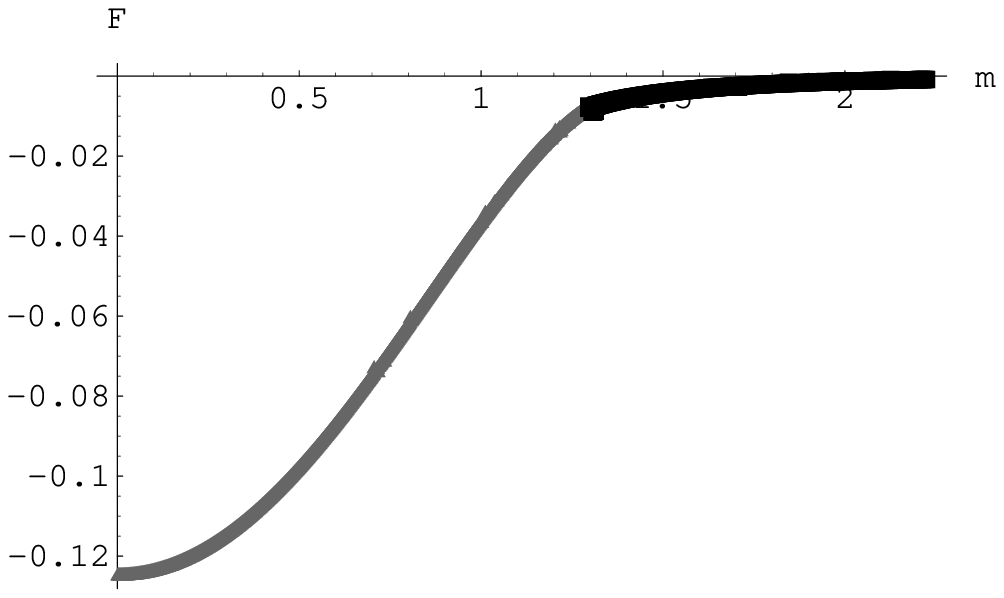} &
\includegraphics[width=0.4\textwidth]{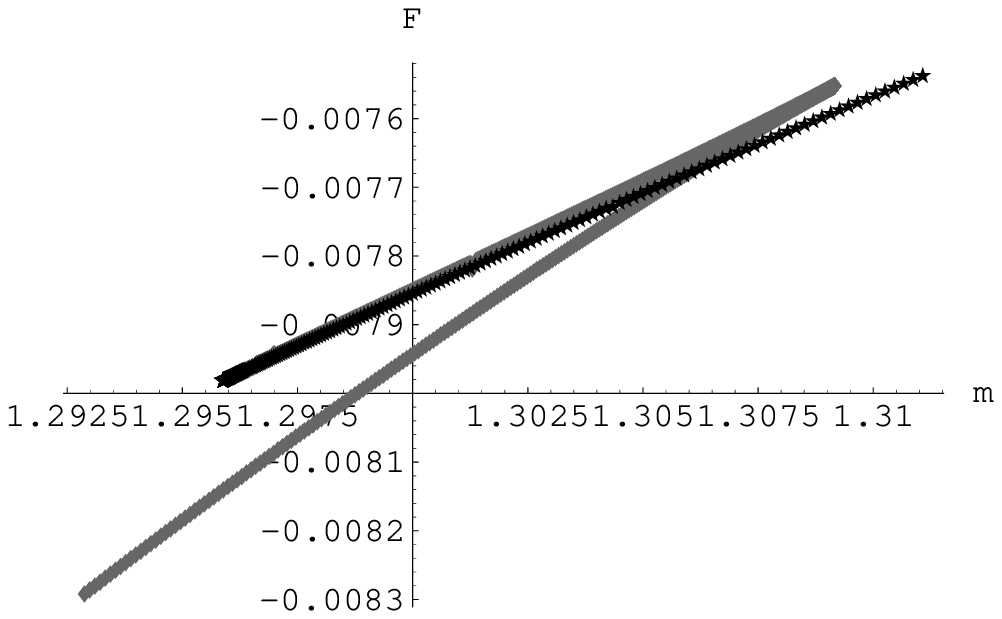}\\
(a) & (b)
\end{tabular}
\caption{\label{flatf}(a.) Free Energy as a function of mass ($M=1$). (b.)
Close-up view of (a.)}
\end{figure}

Our results for the condensate are summarized in figure
(\ref{flatv}) and for the free energy in (\ref{flatf}), clearly reproducing the first order phase transition found in earlier work. The condensate as a function of mass is multivalued; the plot of the free energy shows that at a critical value of the mass we discontinuously jump from a D7 ending some finite distance away from the horizon to a D7 touching the horizon. In terms of the induced metric on the worldvolume of the D7 brane this transition corresponds to a topology-changing phase transition.

\begin{figure}
\center
\includegraphics[width=0.4\textwidth]{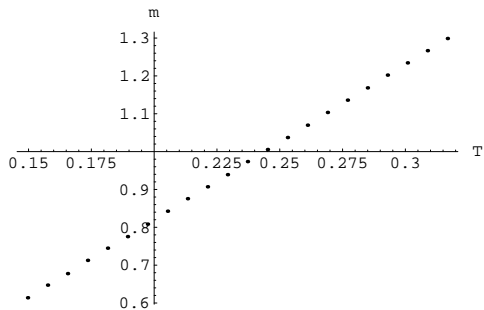}
\caption{\label{flatphase} Phase Diagram of the $\N = 2$ theory in flat space. $T = M / \p$.}
\end{figure}

The phase diagram for this theory appears in figure (\ref{flatphase}). As explained in the introduction, this is a straight line. A linear fit to this data gives a slope of $\approx 4.1$ and a y-intercept of zero. This is expected since in a conformal theory two scales are needed for a transition and either $T$ or $m$ going to zero eliminates one scale.

%%%%%%%%%%%%%%%%%%%%%%%%%%%%%%%%%%%%%%%
\section{The Curved Case}\label{curved}
%%%%%%%%%%%%%%%%%%%%%%%%%%%%%%%%%%%%%%%

\begin{figure}
\center
\begin{tabular}{cc}
\includegraphics[width=0.4\textwidth]{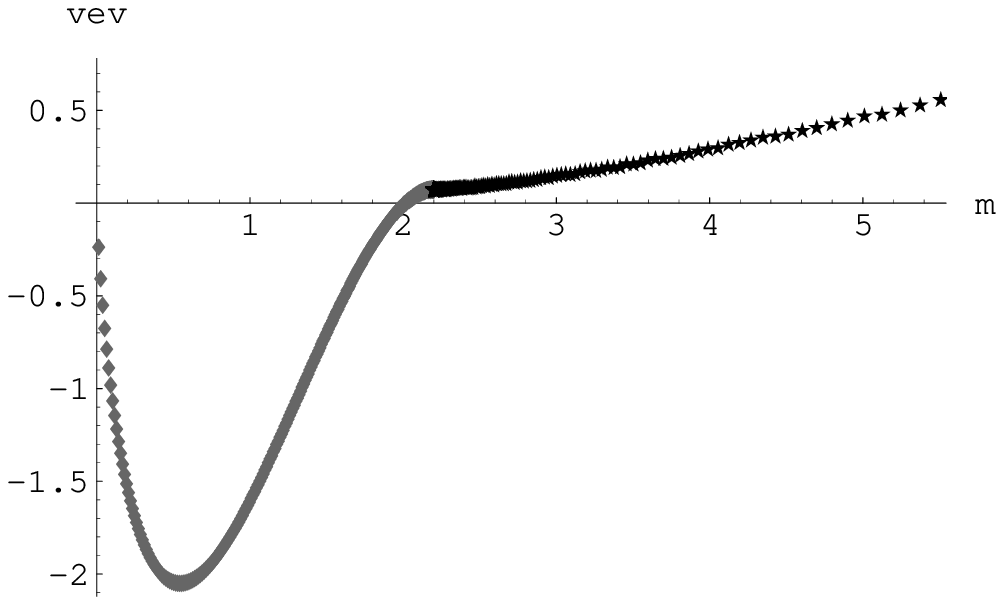} &
\includegraphics[width=0.4\textwidth]{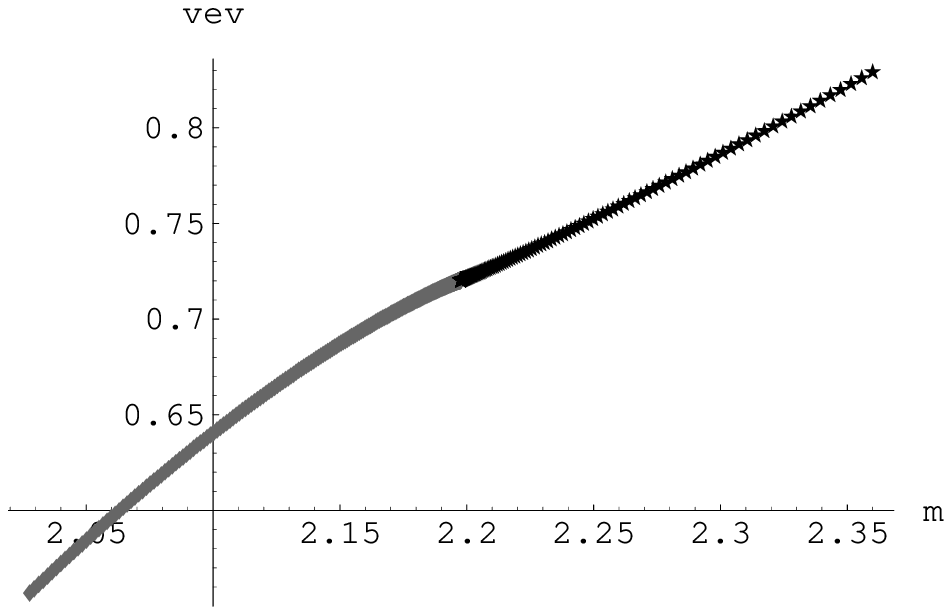}\\
(a) & (b)
\end{tabular}
\caption{\label{vevtads}(a.) Thermal AdS chiral condensate as a function of
mass. Gray curves are branes ending at the center of AdS, black is ending at finite $z<1$ (b.) Close-up of (a.)}
\end{figure}

\begin{figure}
\center
\begin{tabular}{cc}
\includegraphics[width=0.4\textwidth]{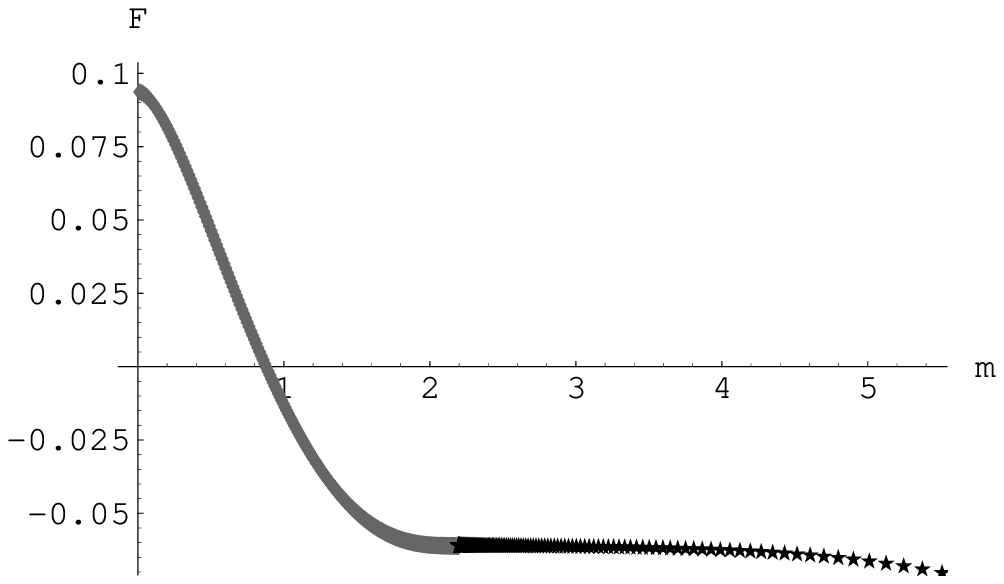} &
\includegraphics[width=0.4\textwidth]{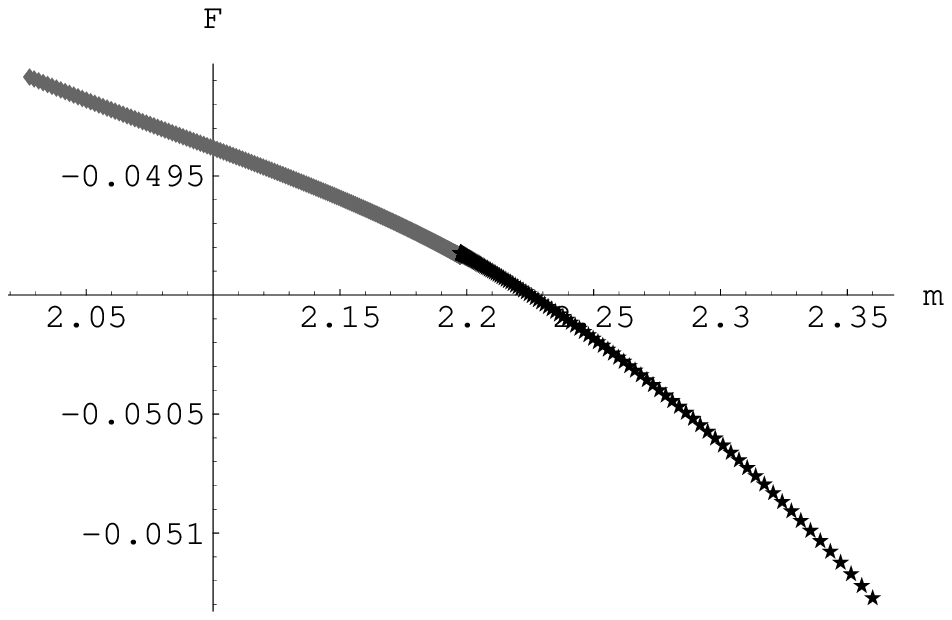}\\
(a) & (b)
\end{tabular}
\caption{\label{ftads}(a.) Thermal AdS free energy as a function of mass. (b.) Close-up of (a.)}
\end{figure}

We start with thermal AdS, for tempratures below $T_{HP}$. We find a surprise. In terms of the topology of the D7 brane there are still two classes of solutions:
the D7 brane can either end at a finite value of the radial coordinate and hence have a finite size $S^3$ inside the AdS$_5$,
or it can reach all the way to the center of AdS$_5$ with this time the $S^3$ inside the $S^5$ remaining finite. The two branches meet at the point where both $S^3$'s shrink at the center of AdS. In the field theory we expect the two configurations to correspond to a competition between the explicit mass terms we added for the hypers with the curvature induced scalar masses. Figure (\ref{vevtads}) shows our results for the vev and figure (\ref{ftads}) shows the free energy. Neither the free energy nor the condensate is multiple-valued in the mass. In other words, despite the topology-changing transition in the bulk we find no first-order transition in the field theory.

Note that our free energy has a non-vanishing positive value of $0.09 \approx 3/32$ at zero temperature. This can be interpreted as the $\lambda N_f N_c$ contribution to the Casimir energy of the system on the sphere. In the massless case the solution can even be obtained analytically, $\theta(z)=0$, and we get an analytic expression for the DBI
contribution to the Casimir energy,

\beq
E_{C,N_c}= \frac{V}{g_5^2} \frac{3}{32} = \frac{ 3 \lambda N_f N_c}{2^9 \pi^4} V
\eeq

where $V$ denotes the volume of the 3-sphere. The leading order (in
$N_c$) contribution to the Casimir energy comes purely from the
supergravity action evaluated on background AdS$_5$ $\times$ $S^5$
geometry itself and has been obtained in
\cite{Balasubramanian:1999re},

\beq
E_{C,N_c^2} = \frac{ 3 N_c^2}{32 \pi^2} V.
\eeq

Their result matches perfectly the free field theory result, where one basically just counts the number of fields
and hence gets a contribution of order $N_c^2$ from the adjoint fields with no powers of $\lambda$. This agreement is guaranteed by supersymmetry. In \cite{Cappelli:1988vw,Skenderis:2000in} it was
shown that the Casimir energy can be obtained from the conformal
anomaly by exploiting the fact that the sphere is conformal to flat
space and the latter is assigned zero Casimir energy. The conformal
anomaly however is completely protected (that is, independent of the
coupling constant) in a theory with at least ${\cal N}=2$
supersymmetry\footnote{With ${\cal N}=1$, supersymmetry still relates
the conformal anomaly to a protected R-current anomaly, but different U(1)s
can play the role of the superconformal R-current at different
values of the coupling.} and hence so is the Casimir energy on the
sphere (or any other space conformal to flat space).

The order $N_f N_c$ correction we found to this leading order
Casimir energy is proportional to $\lambda$ and hence can not possibly
match the free field value, but rather seems to come from a 1-loop
contribution with no higher-loop corrections. Since we have
${\cal N}=2$ supersymmetry the conformal anomaly should
still be protected\footnote{Indeed it was shown in
\cite{Aharony:1999rz} that in a closely related system with an O7 in
addition to 4 probe D7s the supergravity anomaly perfectly
reproduces the field theory anomaly. It is easy to see that without
the orientifold their analysis for a single D7 reproduces the right
value corresponding to a fundamental hyper. In their case the field
theory was exactly conformal (not just to leading order in
$\frac{N_f}{N_c}$) and in the bulk the order $\lambda N_f N_c$
contribution to the Casimir energy canceled between the orientifold and
D7s}. However, this time there is an additional contribution to $\<
T^{\mu}_{\mu} \>$ from the scale anomaly due to the non-vanishing
beta function. In an $\N =2$ theory the beta-function is 1-loop exact and hence
sees only the order $\lambda$ correction even at strong coupling. Our result suggests that the Casimir energy might in a
similar fashion only get a contribution from 1-loop. It would be interesting to see if this can be verified from a weak coupling calculation.

\begin{figure}
\center
\begin{tabular}{cc}
\includegraphics[width=0.4\textwidth]{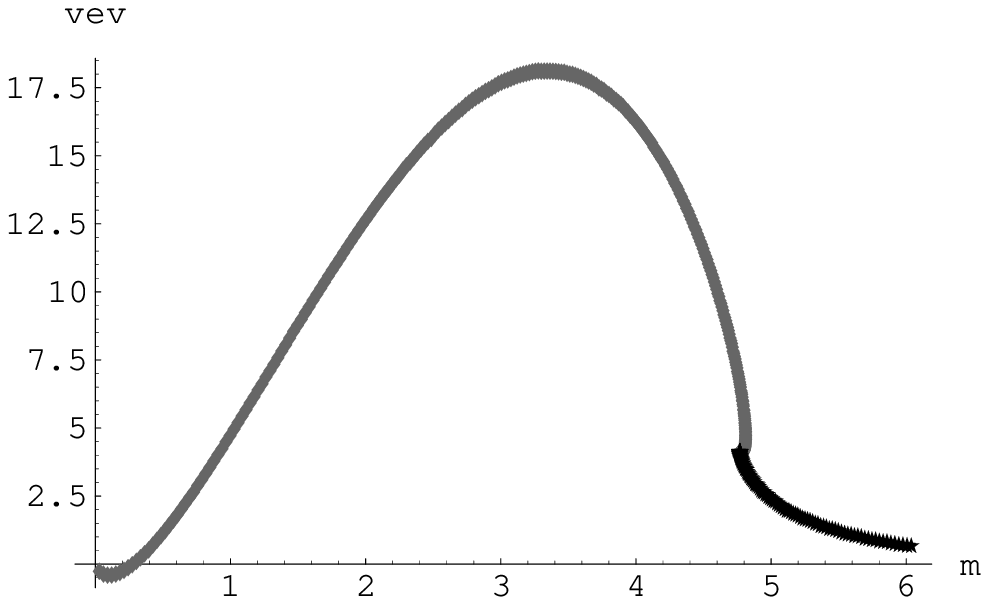} &
\includegraphics[width=0.4\textwidth]{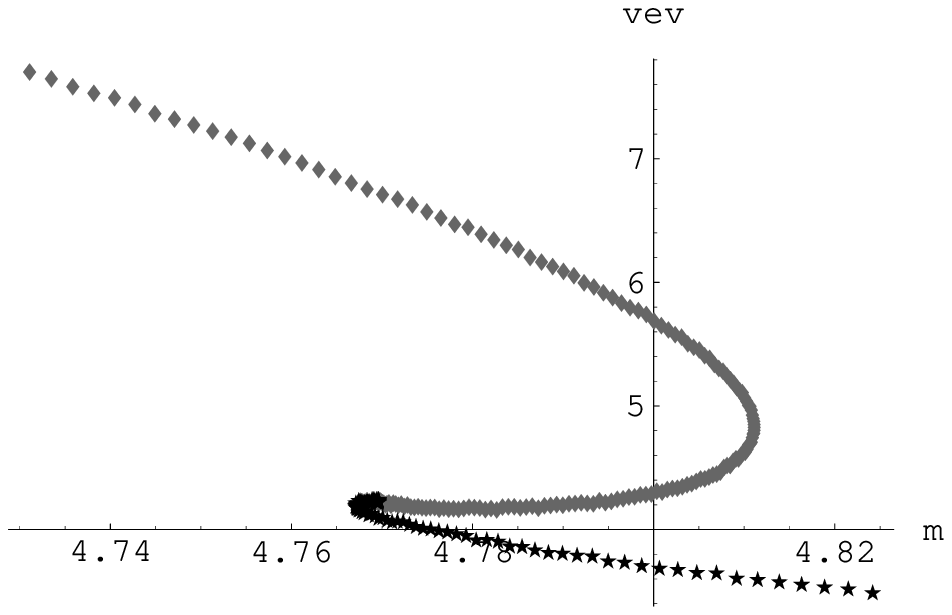}\\
(a) & (b)
\end{tabular}
\caption{\label{vevadsbh}(a.) AdS-Schwarzschild chiral condensate as a function of
mass. Gray curves are branes falling into the horizon, black is ending outside. $T = 1.75 * T_{HP}$ (b.) Close-up of (a.)}
\end{figure}

\begin{figure}
\center
\begin{tabular}{cc}
\includegraphics[width=0.4\textwidth]{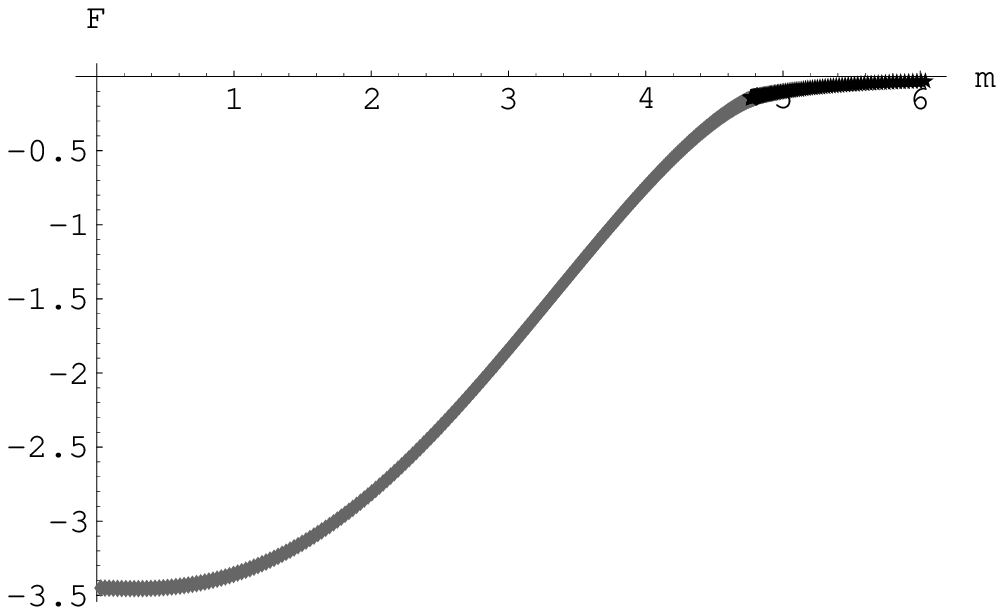} &
\includegraphics[width=0.4\textwidth]{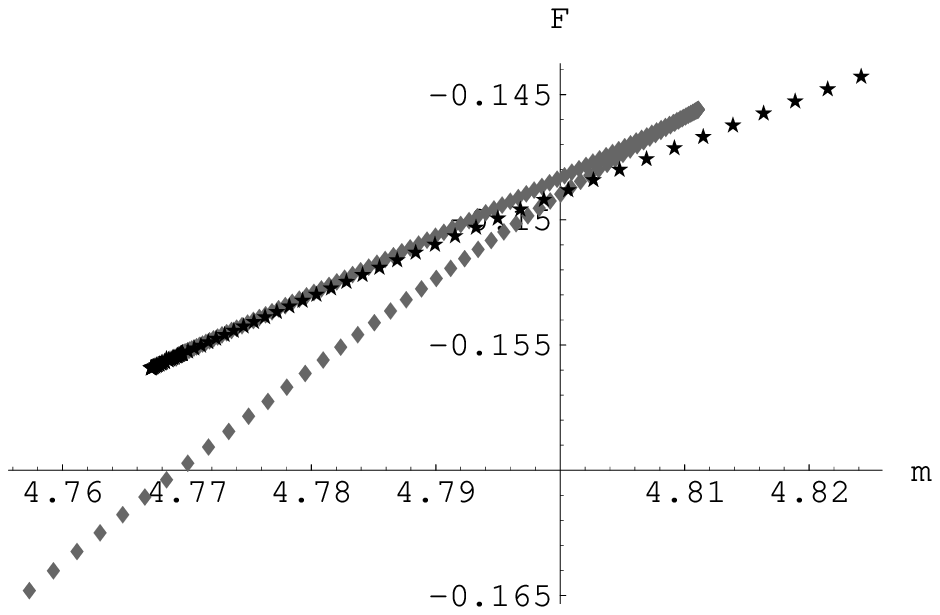}\\
(a) & (b)
\end{tabular}
\caption{\label{fadsbh}(a.) AdS-Schwarzschild free energy as a function of mass. $T = 1.75 * T_{HP}$ (b.) Close-up of (a.)}
\end{figure}

Finally we come to the AdS-Schwarzschild case for $T > T_{HP}$. We find that for $T \gtrsim 4 * T_{HP}$ the free energy and condensate already have nearly the same form as the flat case in figures (\ref{flatv}) and (\ref{flatf}). As $T$ decreases, these continuously deform to the shapes shown in figures (\ref{vevadsbh}) and (\ref{fadsbh}) where $T = 1.75 * T_{HP}$.\footnote{At small mass the condensate begins to dip to negative values around $3.7 * T_{HP}$. It may also dip negative at higher temperatures, but our numerics could not achieve sufficiently high resolution to see this for $T > 3.7 * T_{HP}$.} In close-up we find that the condensate is again double-valued and will jump discontinuously. The free energy again has a discontinuous first derivative just as in the flat case. This remains true all the way down to $T_{HP}$ so no critical point analogous to the point C in the large-$N_c$ QCD phase diagram appears for this theory.

Putting both the high temperature and low temperature regions together we finally arrive at the phase diagram for our theory, as displayed in figure (\ref{phasediagram}). No first-order transition occurs below the $T = T_{HP}$ vertical deconfinement transition line. A linear fit indicates that the chiral transition line has slope $\approx 5.8$. This agrees with the earlier result of $4.1$ from the flat case after taking into account the factor of $\sqrt{2}$ difference between the $z$ coordinate of our flat-case metric and that of our AdS-Schwarzschild metric: $5.8 \approx \sqrt{2} * 4.1$. The chiral transition line intercepts the deconfinement line at $m \approx 2.6$.

%%%%%%%%%%%%%%%%%%%%%%%%%%%%%%%%%%%%%%
\section{Conclusion}\label{conclusion}
%%%%%%%%%%%%%%%%%%%%%%%%%%%%%%%%%%%%%%

We have presented a numerically-generated phase diagram for large-$N_c$ $\mathcal{N}=4$ super-Yang-Mills on a 3-sphere coupled to $N_f$ massive fundamental hypers, including the first-order chiral phase transition line. Below the Hawking-Page temperature we find that, despite a topology-changing transition in the bulk, a first-order transition does not occur in the boundary theory.

Isospin chemical potential could be included in this analysis by introducing multiple D7's ending at different values of the AdS radius, allowing for 7-7 strings and hence heavy-light mesons. Baryon number chemical potential could be included by turning on the gauge potential on the D7 worldvolume, although it is unclear what boundary conditions to impose on this field in the IR. Supposing that question could be answered, this would be one advantage AdS/CFT would have over lattice gauge theory which currently cannot easily include baryon number chemical potential. Other interesting extensions include studying the corresponding phase diagram at weak coupling and studying the effects of the D7 backreaction.

\section*{Acknowledgements}
We would like to thank C.R. Graham, C. Herzog, D. Mateos, S. Sharpe, K. Skenderis, M. Strassler, and L. Yaffe for helpful discussions. A. O'B. would also like to thank D. Yamada. The work of A.K. was supported in part by DOE contract \# DE-FG02-96-ER40956. The work of A. O'B. was supported by a Jack Kent Cooke Foundation scholarship.

\end{document}